\documentclass[12pt]{article}

\usepackage[utf8]{inputenc}
\usepackage[T1]{fontenc}
\usepackage{microtype}
\usepackage{xcolor}
\usepackage{graphicx}
\usepackage{dcolumn}
\usepackage{multirow}
\usepackage{siunitx}
\usepackage[normalem]{ulem}
\DeclareSIUnit\Dq{Dq}
\DeclareSIUnit\bohrmagneton{\ensuremath{\mu_{\mathrm{B}}}}
\DeclareSIUnit\meV{\ensuremath{meV}}
\DeclareSIUnit\angstrom{\text {Å}}
\usepackage{amsmath}

\usepackage{float}
\usepackage[section]{placeins}
\usepackage{braket}

\usepackage[numbers,sort&compress]{natbib}

\setcitestyle{
    numbers,
    open={(},
    close={)},
    citesep={,}
}

\usepackage[
    colorlinks=true,
    citecolor=black,
    linkcolor=black,
    urlcolor=blue
]{hyperref}
\usepackage{lmodern}

\usepackage{listings}
\lstdefinestyle{pystyle}{
  language=python,
  columns=fullflexible,
  keywordstyle={\color[HTML]{228b22}},
  identifierstyle={\color[HTML]{0000ff}},
  commentstyle={\color[HTML]{b22222}},
  stringstyle={\color[HTML]{8b2252}},
  morecomment=[l][{\color[HTML]{483d8b}}]{\#},
}

\newenvironment{sciabstract}{%
\begin{quote} \bf}
{\end{quote}}

\newcounter{lastnote}

\title{Excitonic magnetism and competing valence bond solid instabilities in $d^4$ ruthenium compounds} 

\author
{Aleksandra Krajewska,$^{1*}$ 
Tomohiro Takayama$^{1,2,3*}$\\
and Hidenori Takagi$^{1,2,3,4*}$\\
\\
\normalsize{$^{1}$Max Planck Institute for Solid State Research, Stuttgart, Germany}\\
\normalsize{$^{2}$Institute for Functional Matter and Quantum Technologies, Stuttgart, Germany}\\
\normalsize{$^{3}$National Institute for Materials Science, Tsukuba, Japan}\\
\normalsize{$^{4}$University of Tokyo, Tokyo, Japan}\\
\normalsize{$^\ast$to whom correspondence should be addressed:}\\
\normalsize{$^\ast$e-mail:  a.krajewska@fkf.mpg.de, takayama.tomohiro@nims.go.jp, h.takagi@fkf.mpg.de}
}

\date{}

\begin{document} 


\baselineskip24pt


\maketitle 

\newpage

\begin{sciabstract}
Spin-orbit entangled $d^4$ Mott insulators serve as a rich playground for unconventional magnetism and competing quantum electronic ground states. Intra-ionic spin-orbit coupling $\lambda_{\rm{SOC}}$ generates a non-magnetic $J_{\rm{eff}} = 0$ singlet ground state separated from an excited $J_{\rm{eff}} = 1$ triplet. The effective intra-atomic gap ${\Delta}_{\rm{eff}}$ between the ground state and lowest excited state is determined by $\lambda_{\rm{SOC}}$ and non-cubic crystal field ${\Delta}_{\rm{cf}}$. The intersite magnetic exchange $J_{\rm{ex}}$ induces dispersion of the excited $J_{\rm{eff}} = 1$ triplet. When the triplet bandwidth $W_{\rm{t}}$ overcomes ${\Delta}_{\rm{eff}}$, the system undergoes long-range magnetic ordering which is viewed as a condensation of spin-orbit excitons and called excitonic magnetism. In this review, we systematically survey representative $d^4$ ruthenium compounds across four distinct structural connectivities of Ru-ligand octahedra, from K$_2$RuCl$_6$ with isolated octahedra, corner-sharing perovskite Ca$_2$RuO$_4$, edge-sharing honeycombs, to corner-sharing pyrochlores, and map them onto a conceptual parameter axis of $W_{\rm{t}} / {\Delta}_{\rm{eff}}$. We demonstrate that $d^4$ ruthenium compounds not only represent evolution from a $J_{\rm{eff}} = 0$ Mott state to excitonic magnetism but encompass a broader competition of diverse electronic phases arising from spin-orbit coupling, exchange interactions, electron hopping, crystal field and bond covalency inhabiting similar energy scales. Near excitonic quantum critical point, where $W_{\rm{t}} \sim {\Delta}_{\rm{eff}}$, lattice geometry and chemical bonds of constituent atoms play a decisive role in selecting the electronic ground state, frequently driving valence bond solid formation that pre-empts excitonic quantum criticality.
\end{sciabstract}

\section*{Introduction}

Strong spin-orbit coupling (SOC) in heavy 4$d$/5$d$ transition-metal Mott insulators gives rise to electronic phases distinct from those of 3$d$ systems, where the low-energy degrees of freedom are often well described with spin-only moments~\cite{Khomskii2014,Witczak-Krempa2014,Takayama2021}. This distinction was first exemplified in the layered perovskite iridate Sr$_2$IrO$_4$. Although the spatially extended 5$d$ orbitals are expected to promote metallicity through large bandwidth and reduced Coulomb repulsion $U$, Sr$_2$IrO$_4$ is an antiferromagnetic insulator~\cite{Crawford1994,Cao1998}. The key ingredient driving the insulating state is a relativistic SOC which scales with the atomic number $Z$ as $\lambda_{\rm{SOC}} \propto Z^2$~\cite{LandauLifshitz1981}; from tens of $\SI{}{\meV}$ in 3$d$ compounds, where it is treated as a weak perturbation, to $\SI{0.1}{\eV}$ in 4$d$ and $\SI{0.5}{\eV}$ in 5$d$ compounds. For 4$d$/5$d$ compounds with strong cubic crystal field, i.e. large $\SI{10}{\Dq}$, $d$-electrons are accommodated first into the three-fold degenerate $t_{2g}$ manifold which has an effective orbital angular moment $L_{\rm{eff}} = 1$. SOC entangles spin moment $S$ and $L_{\rm{eff}}$, yielding spin-orbit-entangled $J_{\rm{eff}}$ states~\cite{Abragam1970}. In Sr$_2$IrO$_4$ with 5$d^5$ Ir$^{4+}$ ion, SOC splits the $t_{2g}$ manifold into the lower-lying $J_{\rm{eff}} = 3/2$ and upper-lying $J_{\rm{eff}} = 1/2$ states. The resulting half-filled $J_{\rm{eff}} = 1/2$ band is narrow enough for a moderate $U$ to open a charge gap, leading to a spin-orbit-entangled Mott insulating state~\cite{Kim2008,Kim2009}.

In such spin-orbit-entangled Mott insulators, the spin-orbital wavefunctions inherit directional character of orbital interactions and therefore the inter-site magnetic exchange depends sensitively on the bond geometry~\cite{Khaliullin2005}. In the case of $d^5$ $J_{\rm{eff}} = 1/2$ systems, electron hopping across the $\SI{180}{\degree}$ Ir--O--Ir corner-sharing bond conserves the spin and orbital components, namely conserves $J_{\rm{eff}} = 1/2$ pseudospin. This process leads to an isotropic antiferromagnetic Heisenberg exchange~\cite{Jackeli2009}, which has been identified in Sr$_2$IrO$_4$~\cite{Fujiyama2012}. In contrast, across the edge sharing $\SI{90}{\degree}$ Ir--O$_2$--Ir bonds, hopping through ligand oxygen atom does not conserve the orbital component, and this process fully suppresses conventional Heisenberg exchange. Instead, the higher-order process involving $J_{\rm{eff}} = 3/2$ manifold yields bond-dependent Ising interaction, which is dubbed Kitaev coupling~\cite{Jackeli2009}. The distinct exchange interactions between $\SI{180}{\degree}$ and $\SI{90}{\degree}$ bonding geometry contrast with those of the spin-only $S = 1/2$ systems. 

Whereas the $d^5$ iridates provide a natural entry point to spin-orbit-entangled magnetism, the local wavefunctions and their exchange interactions change markedly depending on the $d$-electron configuration. In $d^1$ and $d^2$ systems, SOC produces $J_{\rm{eff}} = 3/2$ and $J_{\rm{eff}} = 2$ ground state, respectively. The $J_{\rm{eff}} = 3/2$ and 2 states host higher-order multipoles such as charge quadrupoles and magnetic octupoles and the emergence of multipolar physics has been proposed~\cite{Chen2010,Paramektani2020,Khalliulin2021}. The double-perovskite oxides containing 5$d$ elements emerged as possible realization of multipolar ordering~\cite{Hirai2020,Maharaj2020}.

\begin{figure*}[t]
	\centering
	\includegraphics[width=0.65\linewidth,trim=0 790 880 0,clip]{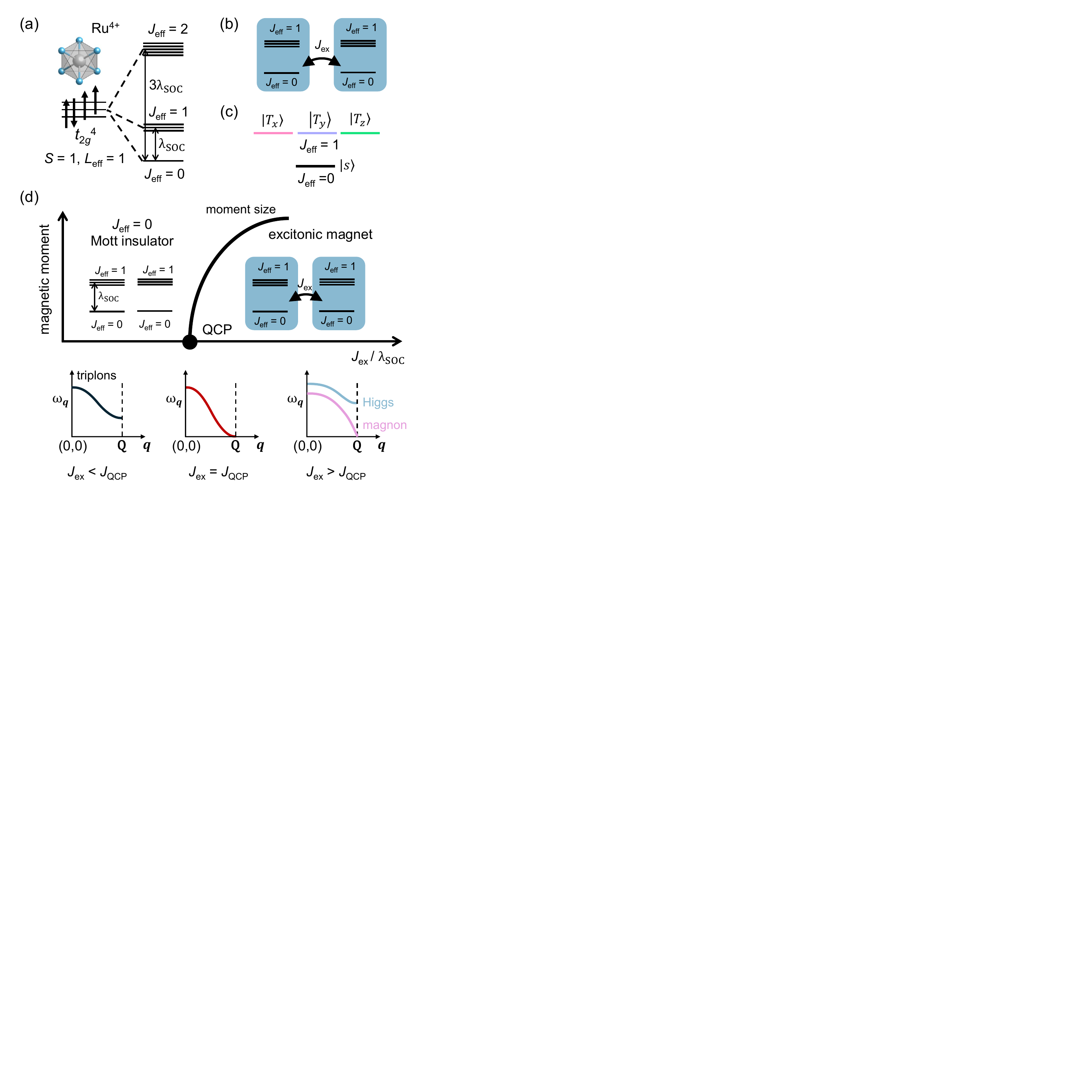}
    \caption{\textbf{Concept of excitonic magnetism for $d^4$ electronic configuration.} (a) Energy level scheme of a $t_{2g}^4$ configuration in the $LS$ coupling regime, where $J_{\rm{eff}} = 0$ singlet, $J_{\rm{eff}} = 1$ triplet and $J_{\rm{eff}} = 2$ quintet are formed with $\lambda_{\rm{SOC}}$. (b) Inter-site exchange interactions $J_{\rm{ex}}$ involving triplon hopping or triplon pair creation and annihilation. (c) $J_{\rm{eff}} = 0$ expressed as singlet $\ket{s}$ and $J_{\rm{eff}} = 1$ triplet expressed as three triplons $\ket{T_i}$. (d) Top: Ordered moment of an excitonic magnet depending on the ratio of $J_{\rm{ex}}$ to the spin-orbit coupling $\lambda_{\rm{SOC}}$. The size of the ordered moment changes depending on the degree of singlet-triplet admixture. Quantum critical point (QCP) separates the $J_{\rm{eff}} = 0$ Mott insulator and excitonic magnet regimes. Bottom: The dispersion relation of elementary excitations with energy ${\omega}_q$ as a function of momentum $q$. The triplon excitations acquire dispersion by $J_{\rm{ex}}$ and soften at an ordering wavevector $\mathbf{Q}$. For critical $J_{\rm{ex}}$ value, $J_{\rm{QCP}}$, the triplon bandwidth crosses the singlet. For $J_{\rm{ex}} > J_{\rm{QCP}}$ in the excitonic magnet phase, the three triplon bands split into a doubly degenerate gapless magnon and gapped longitudinal Higgs mode.}
	\label{fig:concept}
\end{figure*}

$d^4$ configuration provides a unique situation; SOC stabilises a non-magnetic $J_{\rm{eff}} = 0$ singlet ground state out of $S = 1$ and $L_{\rm{eff}} = 1$~\cite{Abragam1970}. In such $J_{\rm{eff}} = 0$ Mott insulators, despite a local non-magnetic singlet state, magnetic ordering may take place as a result of inter-site magnetic exchange $J_{\rm{ex}}$. The $J_{\rm{eff}} = 0$ singlet state has a $J_{\rm{eff}} = 1$ triplet as the lowest excited state, where their gap is imposed by SOC. When inter-site exchange interactions overcome the local spin-orbit gap, magnetic ordering occurs, which is viewed as a condensation of the excited triplet state, namely, spin-orbit excitons, and thus called excitonic magnetism~\cite{Khaliullin2013}. Hence, the spin-orbital ground state of $d^4$ system is determined by the balance of spin-orbit coupling $\lambda_{\rm{SOC}}$ and exchange interaction $J_{\rm{ex}}$; a non-magnetic singlet state is formed when $J_{\rm{ex}}$ is small compared to the spin-orbit gap, while long-range magnetic order appears with sufficiently strong $J_{\rm{ex}}$~\cite{Svoboda2017}. The two spin-orbital ground states are separated by a quantum critical point (QCP) where $\lambda_{\rm{SOC}}$ and $J_{\rm{ex}}$ are comparable. 

This review focuses on 4$d^4$ ruthenium compounds which appeared as a promising platform for excitonic magnetism. In Ru$^{4+}$ compounds the spin-orbit coupling and the inter-site exchange fall within the same $\SIrange{10}{100}{\meV}$ energy window, enabling the access to the spin-orbital ground states ranging from $J_{\rm{eff}} = 0$ singlet to excitonic magnet. Indeed, as discussed below, the two spin-orbital ground states have been identified in ruthenium compounds. However, other competing electronic ground states, such as valence bond solid state, have been often observed, which may reside in the vicinity of the excitonic quantum critical point. This may indicate subtle balance of electronic parameters near the QCP, including not only spin-orbit coupling and exchange interactions, but also electron hopping, non-cubic crystal field, and lattice energy, which give rise to rich spin-orbital phases. We first introduce the single-ion and exchange physics of excitonic magnetism. We then survey the representative $d^4$ ruthenium compounds organized by RuO$_6$ (RuCl$_6$) octahedra connectivity and argue that a valence bond instability intervenes near the critical point. We finally summarize the broad range of ruthenium compounds in terms of the single parameter, the ratio of the bandwidth of the dispersive triplet and the local spin-orbit gap, and discuss some future perspectives.

\section*{Excitonic magnetism of spin-orbit-entangled $d^4$ system}

\begin{figure*}[t]
	\centering
	\includegraphics[width=0.9\linewidth,trim=100 360 210 0,clip]{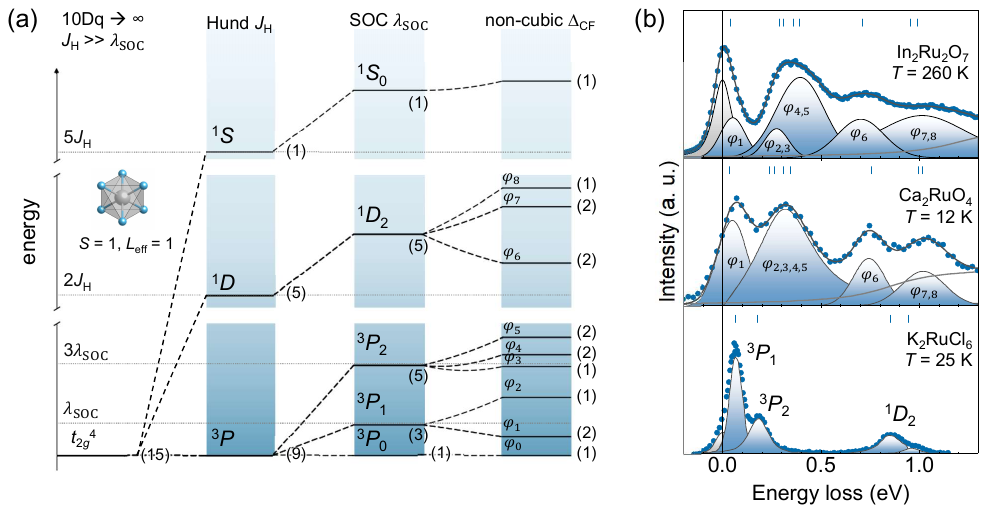}
	\caption{\textbf{Single-ion physics for $d^4$ configuration.} (a) $t_{2g}^{4}$ energy level diagram in the $LS$ coupling scheme in octahedral crystal field as a function of Hund’s coupling $J_{\rm{H}}$, spin-orbit coupling $\lambda_{\rm{SOC}}$ and positive non-cubic crystal field ${\Delta}_{\rm{cf}}$, assuming $\SI{10}{\Dq} \rightarrow \infty$ as well as $J_{\rm{H}} >> \lambda_{\rm{SOC}}$. Numbers in brackets indicate level multiplicity. (b) Resonant inelastic x-ray scattering (RIXS) spectra of several ruthenium compounds collected at Ru $L_3$-edge. Fitted peaks are  ascribed to the excitations to higher-energy spin-orbital multiplets shown in (a) and marked with blue ticks. The grey line represents the electronic continuum. Fitted parameters are listed in the main text. Spectra were taken at $(H,K,L)$ of (0.51, 0.51, -0.32) and (0.32, 0.32, 0.28) r.l.u. for K$_2$RuCl$_6$ and Ca$_2$RuO$_4$, respectively, while In$_2$Ru$_2$O$_7$ was measured in powder form. RIXS spectra are reproduced from Ref.~\cite{Krajewska2024,Gretarsson2019,Takahashi2021}.}
	\label{fig:RIXS}
\end{figure*}

For a $d^4$ configuration in a strong octahedral crystal field, the four $d$-electrons occupy the low-lying $t_{2g}$ orbitals in the low-spin configuration, yielding a $S = 1$ and $L_{\rm{eff}} = 1$ manifold (Fig.~\ref{fig:concept}(a)). With the $e_g$ orbitals projected out and in the $LS$ coupling limit, Hund’s rules define ${^{3}P}$ ($S = 1$, $L_{\rm{eff}} = 1$) configuration as the ground state, the ${^{1}D}$ ($S = 0, L_{\rm eff} = 2$) configuration as the next excited state and the ${^{1}S}$ ($S = 0, L_{\rm eff} = 0$) configuration as the highest energy state, as shown in Fig.~\ref{fig:RIXS}(a). SOC entangles the spin and orbital degrees of freedom and splits the lowest energy ${^{3}P}$ multiplet into a $J_{\rm{eff}} = 0$ singlet ground state, excited $J_{\rm{eff}} = 1$ triplet and higher-energy $J_{\rm{eff}} = 2$ quintet. The gap between $J_{\rm{eff}} = 0$ and $J_{\rm{eff}} = 1$ state corresponds to $\lambda_{\rm{SOC}} = {\zeta}/2S = {\zeta}/2$ where $\zeta$ represents the atomic spin-orbit coupling. The ground state is nominally non-magnetic $J_{\rm{eff}} = 0$ state, carrying no dipole moment at the single-ion level.

The hierarchy of the singlet ground state and triplet excited state is analogous to quantum dimer systems such as TlCuCl$_3$, where antiferromagnetically coupled $S = 1/2$ spins on Cu pairs form a $S = 0$ singlet separated from an excited $S = 1$ triplet~\cite{Nikuni2000,Ruegg2003,Giamarchi2008}. The key differences, however, lie in the nature of the singlet state, its temperature-dependent magnetic response and the primary mechanism driving the triplet condensation.

In quantum spin dimers the gap is scaled by the intra-dimer magnetic exchange. The spin singlet carries no magnetic dipole moment and has no off-diagonal matrix elements with the triplet under external magnetic field. At low temperatures the magnetic susceptibility drops exponentially to zero and magnetic response emerges only at high temperatures due to the thermal population of the triplet. External magnetic field acts via Zeeman splitting and linearly splits the triplet, lowering the $S^{z} = +1$ component until it crosses the singlet ground state and condenses at a critical field $H_c$, leading to a Bose-Einstein condensation of the triplet state.

In contrast, the singlet-triplet gap in the $d^4$ case is set by the intra-ionic SOC. Although the local $J_{\rm{eff}} = 0$ carries no magnetic dipole moment, the magnetisation $\mathbf{M} = \mu_{\text{B}}(\mathbf{L}_{\text{eff}} + 2\mathbf{S})$ possesses finite off-diagonal matrix elements between the $J_{\rm{eff}} = 0$ singlet and $J_{\rm{eff}} = 1$ triplet. An external magnetic field therefore induces mixing between the singlet and triplet states, polarising the ground state wavefunction to yield a finite, temperature-independent van Vleck susceptibility at low temperatures~\cite{VanVleck1932}. With increasing temperature, thermal population of the $J_{\rm{eff}} = 1$ manifold gives rise to a Curie-like component. This off-diagonal mixing induces singlet-triplet level repulsion, rendering external magnetic field ineffective for driving triplon condensation.

Instead, the primary driving force for triplon condensation, namely long-range magnetic order, is the inter-site magnetic exchange $J_{\rm{ex}}$~\cite{Khaliullin2013,Svoboda2017} (Fig.~\ref{fig:concept}(b)). The collective magnetism is described by representing $J_{\rm{eff}} = 0 \rightarrow 1$ excitations as triplons $T_i$, expressed in Cartesian flavours $T_x$, $T_y$ and $T_z$, as shown in Fig.~\ref{fig:concept}(c). For a weak $J_{\rm{ex}}$, much smaller than the singlet-triplet gap, the ground state remains predominantly the $J_{\rm{eff}} = 0$-like singlet state. By increasing $J_{\rm{ex}}$, the inter-site exchange translates into triplon hopping and generates triplon dispersion across the Brillouin zone, causing the triplons to soften at the magnetic ordering vector $\mathbf{Q}$. Eventually, when $J_{\rm{ex}}$ reaches a critical value and the triplon bandwidth $W_{\rm{t}}$ overcomes the gap, the triplons undergo Bose-Einstein condensation, giving rise to long-range magnetic order called excitonic magnetism (Fig.~\ref{fig:concept}(d)). This contrasts with the spin-$1/2$ dimer systems where inter-dimer interaction, which would give rise to dispersion of the triplet, is generally much smaller than the intra-dimer exchange and spontaneous triplet condensation without the application of magnetic field rarely occurs~\cite{Ruegg2004}.

The ordered magnetic moment is determined by the triplon condensate density, namely, the degree of singlet and triplet admixture. In the ordered state, the triplon excitations split into a doubly degenerate gapless Goldstone magnon mode and a gapped amplitude mode (Fig.~\ref{fig:concept}(d)). This amplitude mode, which represents fluctuation of the moment size, is the magnetic analogue of the Higgs mode. Observation of this longitudinal mode distinguishes an excitonic magnet from a conventional magnet with rigid local spin moments. Furthermore, since the $J_{\rm{eff}} = 1$ triplet states inherit the orbital moment of $t_{2g}$ manifold, triplon hopping becomes flavour- and bond-dependent, giving rise to highly anisotropic magnetic exchange and frustrated excitonic magnetism which is discussed below for the $d^4$ honeycomb compounds.

In most real materials, the connected transition-metal-ligand octahedra deviate from the perfect cubic environment and feature tetragonal, orthorhombic or trigonal distortions, leading to a significant non-cubic crystal field ${\Delta}_{\rm{cf}}$ of the order of $\SIrange{50}{200}{\meV}$ which lifts the degeneracy of excited $J_{\rm{eff}} = 1$ and $2$ states. Therefore, the effective spin-orbital gap ${\Delta}_{\rm{eff}}$ is governed not only by SOC, but also by ${\Delta}_{\rm{cf}}$. For the case of tetragonal or trigonal distortion, the $J_{\rm{eff}} = 1$ is split into a singlet and a doublet, and, depending on the sign of the distortion, either of the components is brought closer towards the $J_{\rm{eff}} = 0$ singlet (Fig.~\ref{fig:RIXS}(a)). This distortion therefore does not simply quench the orbital moments as assumed in the spin-only picture but plays a constructive role for excitonic magnetism by reducing ${\Delta}_{\rm{eff}}$. On the other hand, $W_{\rm{t}}$ is governed by $J_{\rm{ex}}$ and the coordination number, where $W_{\rm{t}}$ increases with the number of neighbours. Competing exchange interactions can lead to frustration, which may reduce the triplon dispersion and thereby suppress long-range magnetic order. The competition between $J_{\rm{eff}} = 0$ Mott insulator and excitonic magnet is therefore defined by the ratio of the effective spin-orbit gap ${\Delta}_{\rm{eff}}$ and the triplon bandwidth $W_{\rm{t}}$, where ${\Delta}_{\rm{eff}}$ is controlled by $\lambda_{\rm{SOC}}$ and ${\Delta}_{\rm{cf}}$, and $W_{\rm{t}}$ is governed by the magnitude and nature of the magnetic exchange $J_{\rm{ex}}$ as well as coordination number.

The Bose-Einstein condensation of spin-orbit excitons is most robust in the large $W_{\rm{t}}$ limit where the ordered moment is large. Near the QCP the ordered moment is small, the condensate is strongly fluctuating, and the Higgs mode softens and carries increasing spectral weight. Exotic spin-orbital phases might emerge as a result of strong quantum fluctuations. However, as we argue below, the system also becomes more susceptible to competing instabilities near the QCP and other electronic phases, such as valence-bond solid or molecular-orbital formation, frequently pre-empt the excitonic QCP. The real $d^4$ materials therefore display even richer spin-orbital phases than the canonical excitonic magnetism picture.

\section*{Materials landscape}

The experimental realization of the spin-orbit singlet-triplet physics depends on the hierarchy of competing energy scales. $d^4$ systems with strong SOC, such as 5$d$ iridates and osmates featuring Ir$^{5+}$ and Os$^{4+}$ appear as prime candidates. While the large SOC stabilizes a robust $J_{\rm{eff}} = 0$, the singlet-triplet gap is of the order of several hundred $\SI{}{\meV}$, as observed in double perovskites $A_2$$B$IrO$_6$ ($A$ = Ba, Sr, $B$ = Lu, Sc)~\cite{Aczel2022}, which is too large to be overcome by realistic inter-site exchange. Ruthenium compounds featuring RuO$_6$ (RuCl$_6$) octahedra with Ru$^{4+}$ cations and $d^4$ configuration provide a more suitable platform in this respect, as all the relevant parameters take comparable values: $\lambda_{\rm{SOC}} \sim \SI{80}{\meV}$, Hund’s coupling $J_{\rm{H}} \sim \SI{0.4}{\eV}$, non-cubic crystal field splitting ${\Delta}_{\rm{cf}} \sim \SIrange{50}{200}{\meV}$ and $J_{\rm{ex}} \sim \SIrange{10}{100}{\meV}$~\cite{Abragam1970}. This places ruthenium compounds near the boundary between the non-magnetic $J_{\rm{eff}} = 0$ Mott insulator and excitonic magnet.

The role of SOC in ruthenium compounds was previously underestimated. Structural distortions of the RuO$_6$ octahedra and deviations from ideal cubic symmetry were often assumed to fully quench the orbital moment, motivating spin-only $S = 1$ descriptions~\cite{Braden1998,Taira2002}. This assumption was in fact contradicted by several experimental observations: van Vleck-type magnetic susceptibility in K$_2$RuCl$_6$~\cite{Kamimura1960}, contribution of orbital angular momentum identified in the spin-resolved photoemission spectra of Ca$_2$RuO$_4$~\cite{Mizokawa2001}, and M{\"o}ssbauer spectra in Y$_2$Ru$_2$O$_7$ described by a spin-orbit-split energy level scheme~\cite{Kmiec2006}. More recently, a consistent picture has emerged from spectroscopic probes. Local~\cite{Das2018,Gretarsson2019,Takahashi2021} and collective excitations~\cite{Jain2017} in ruthenium compounds such as Ca$_2$RuO$_4$ and K$_2$RuCl$_6$ have firmly established SOC as an essential ingredient for understanding their electronic and magnetic properties. The following discussion is organized by connectivity of RuO$_6$ (RuCl$_6$) octahedra, proceeding from the isolated octahedron limit, through corner-sharing perovskite that realises excitonic magnet. We then discuss the systems with $\SI{90}{\degree}$ Ru--O--Ru bonds, focusing on the honeycomb ruthenates, where frustrated excitonic magnetism is anticipated. Finally, we visit the pyrochlore ruthenates comprising a corner-sharing bond with an intermediate Ru--O--Ru angle.

\subsubsection*{K$_2$RuCl$_6$: the single-ion $J_{\rm{eff}} = 0$ limit}

\begin{figure*}[t]
	\centering
	\includegraphics[width=0.6\linewidth,trim=0 1000 550 0,clip]{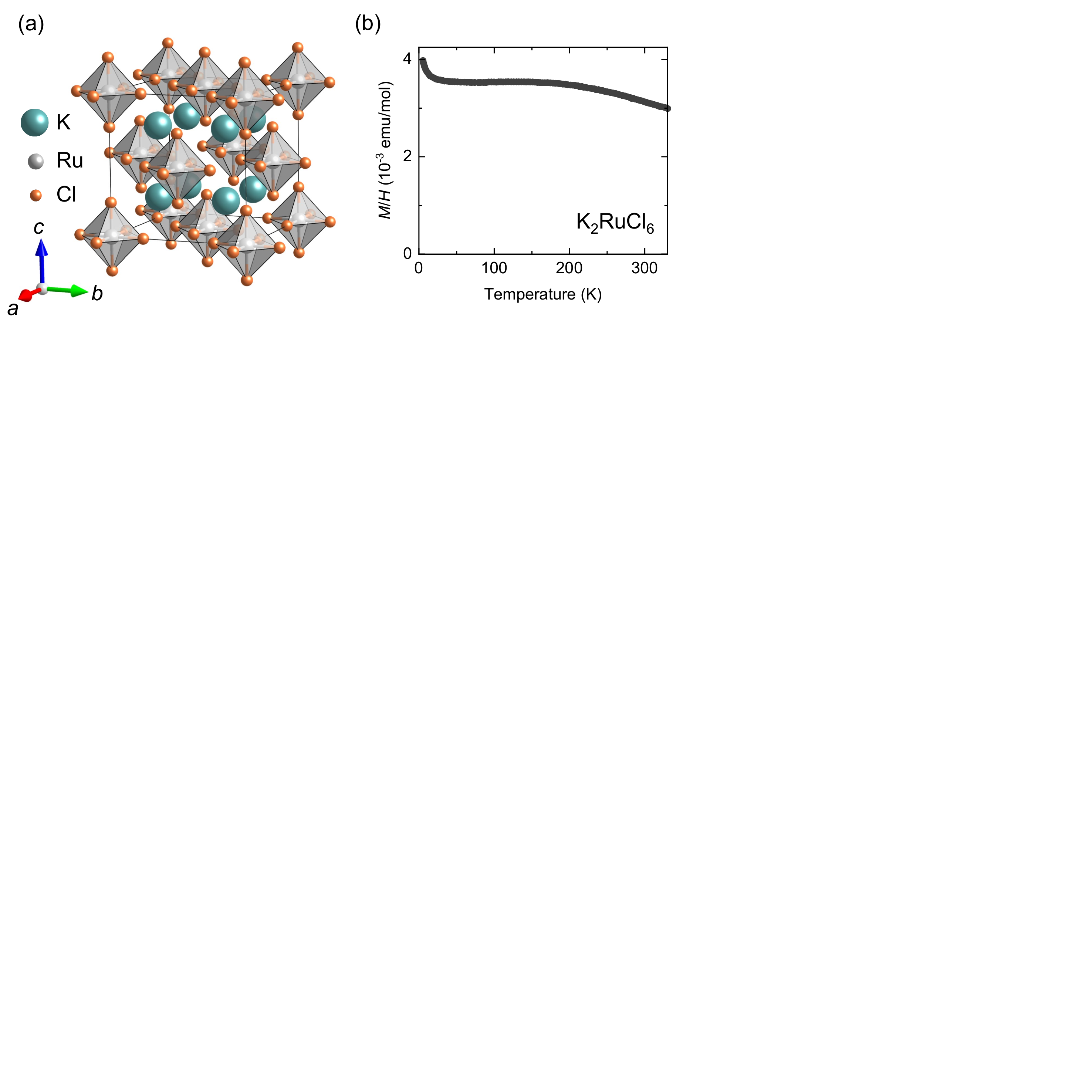}
	\caption{\textbf{K$_2$RuCl$_6$ with a $J_{\rm{eff}} = 0$ singlet state.} (a) Crystal structure of K$_2$RuCl$_6$ ($Fm\overline{3}m$, no. 225). Teal, grey and orange spheres are K, Ru and Cl atoms, respectively. The isolated RuCl$_6$ units form an FCC lattice. (b) Temperature dependence of the magnetic susceptibility of K$_2$RuCl$_6$, showing leveling-off below $\SI{200}{\kelvin}$, a characteristic feature of the van Vleck paramagnetic response. Reproduced from Ref.~\cite{Takahashi2021}.}
	\label{fig:K2RuCl6}
\end{figure*}

To establish $d^4$ ruthenium compounds as a viable platform for exploring excitonic magnetism, it is instructive to first consider the limit where inter-site exchange is minimized ($J_{\rm{ex}} \rightarrow 0$), isolating the underlying single-ion physics. This condition is met in the antifluorite compound K$_2$RuCl$_6$, where RuCl$_6$ octahedra form a face-centered cubic (FCC) lattice but remain spatially well-separated, with Ru--Ru distances near $\SI{7}{\angstrom}$ and no shared ligands~\cite{Adams1952} (Fig.~\ref{fig:K2RuCl6}(a)). The RuCl$_6$ octahedra retain cubic point symmetry and are free from distortions (${\Delta}_{\rm{cf}} = 0$), making K$_2$RuCl$_6$ an ideal model system to probe the $J_{\rm{eff}} = 0$ single-ion limit.

Resonant inelastic X-ray scattering (RIXS) at the Ru $L_3$-edge identifies the local $d^4$ multiplet structure of K$_2$RuCl$_6$ directly~\cite{Takahashi2021} (Fig.~\ref{fig:RIXS}(b)). Transitions from the $J_{\rm{eff}} = 0$ ground state to the $J_{\rm{eff}} = 1$ and $J_{\rm{eff}} = 2$ excited spin-orbital levels below $\SI{0.5}{\eV}$, together with upper-lying Hund multiplets, are clearly resolved and reproduced by a single-ion model with $\lambda_{\rm{SOC}} \sim \SI{50}{\meV}$ and $J_{\rm{H}} \sim \SI{0.42}{\eV}$. Consistent with this assignment, the magnetic response is dominated by van Vleck paramagnetism and no magnetic order is seen due to weak inter-site magnetic interactions, see Fig.~\ref{fig:K2RuCl6}(b). K$_2$RuCl$_6$ thus provides an experimental reference for the $J_{\rm{eff}} = 0$ single-ion limit, comprising Ru$^{4+}$ building block of spin-orbit entangled $d^4$ systems. We note that the reduction of the $\lambda_{\rm{SOC}}$ from the free-ion value of $\sim\SI{80}{\meV}$ may be ascribed to dynamical Jahn-Teller coupling which renormalizes the energy levels of spin-orbital multiplets~\cite{Takahashi2021}.

\subsubsection*{Ca$_2$RuO$_4$: excitonic magnet with corner-sharing RuO$_6$ octahedra}

\begin{figure*}[t]
	\centering
	\includegraphics[width=0.75\linewidth,trim=0 980 300 0,clip]{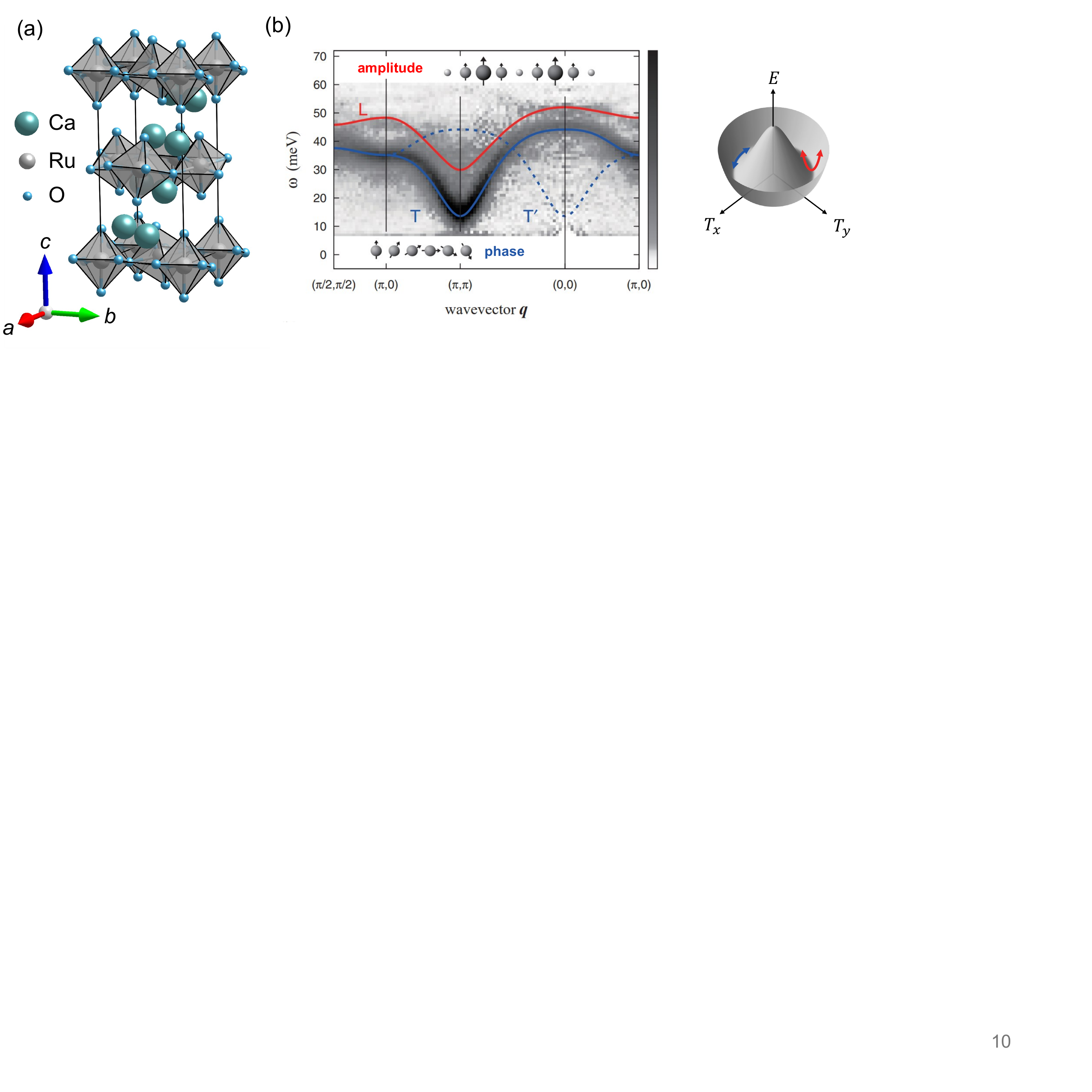}
	\caption{\textbf{Excitonic magnet Ca$_2$RuO$_4$.} (a) Crystal structure of layered perovskite Ca$_2$RuO$_4$ in the orthorhombic phase below $\sim \SI{360}{\kelvin}$ ($Pbca$, no. 61). Teal, grey and blue spheres are Ca, Ru and O atoms, respectively. (b) Inelastic neutron scattering spectrum of Ca$_2$RuO$_4$ below magnetic ordering temperature. The two transverse magnon modes $T$ and $T'$ are marked with a blue solid and dashed line, respectively, and are gapped due to orthorhombic distortion. The longitudinal Higgs mode is marked in red. The phase and amplitude fluctuations are depicted in above and below the spectrum as well as in the Mexican hat potential on the right. Left of (b) is reproduced from Ref.~\cite{Jain2017,Takayama2021}.}
	\label{fig:Ca2RuO4}
\end{figure*}

While K$_2$RuCl$_6$ realises the isolated $J_{\rm{eff}} = 0$ state, the evolution of this ground state under significant $J_{\rm{ex}}$ was first examined in systems with corner-sharing RuO$_6$ octahedra with nearly $\SI{180}{\degree}$ Ru--O--Ru bonds. The layered perovskite Ca$_2$RuO$_4$, which features such a bonding geometry, undergoes a metal-insulator transition at $\sim\SI{360}{\kelvin}$, accompanied by a tetragonal-to-orthorhombic distortion that compresses (along $c$-axis), tilts and rotates the RuO$_6$ octahedra~\cite{Braden1998,Friedt2001} (Fig.~\ref{fig:Ca2RuO4}(a)). The tilt and rotation of the RuO$_6$ octahedra bend the Ru--O--Ru bond from $\SI{180}{\degree}$ to $\sim\SI{150}{\degree}$. Below $T_{\rm{N}} \sim \SI{110}{\kelvin}$, Ca$_2$RuO$_4$ develops long-range antiferromagnetic order~\cite{Nakatsuji1997}.

While the pronounced octahedral distortions were interpreted as support for a spin-only $S = 1$ description~\cite{Anisimov2002,Fang2004,Liebsch2007,Gorelov2010}, Ru $L_3$-edge RIXS has shown spin-orbit coupling to be essential for understanding the low-energy physics of Ca$_2$RuO$_4$~\cite{Gretarsson2019}. Moreover, the non-cubic crystal field splitting ${\Delta}_{\rm{cf}}$ plays a constructive rather than a purely destructive role for excitonic magnetism. The RIXS spectrum resolves a series of intra-$t_{2g}$ excitations reproduced by a single-ion model with $\lambda_{\rm{SOC}} \sim \SI{65}{\meV}$, ${\Delta}_{\rm{cf}} \sim \SI{0.25}{\eV}$ and $J_{\rm{H}} \sim \SI{0.34}{\eV}$, see Fig.~\ref{fig:RIXS}(b). The tetragonal crystal field splits the $J_{\rm{eff}} = 1$ triplet into a lower-lying doublet and an upper-lying singlet, reducing the gap between the ground state singlet to the excited state to approximately $\SI{30}{\meV}$, within the reach of $W_{\rm{t}}$ to drive excitonic condensation.

The magnetic excitation spectrum observed by inelastic neutron scattering reveals the realization of excitonic magnetism~\cite{Jain2017} as shown in Fig.~\ref{fig:Ca2RuO4}(b). The transverse magnon mode T displays a maximum at $\mathbf{q} = (0, 0)$, distinct from that of isotropic Heisenberg antiferromagnetism. This dispersion is consistent with the $XY$-type magnetic anisotropy originating from the local pseudospin-1 state consisting of the singlet and the doublet from the $J_{\rm{eff}} = 1$ state. In addition to the transverse magnon modes, a longitudinal amplitude mode, namely the magnetic Higgs mode, was observed, which is a hallmark of soft-spin, excitonic magnetism. Raman scattering supports the picture of excitonic magnetism through observation of the longitudinal mode in the fully symmetric $A_g$ channel, where contamination from transverse magnons is suppressed~\cite{Souliou2017}. Ca$_2$RuO$_4$ thus constitutes the benchmark realization of an excitonic magnet.

\subsubsection*{Honeycomb ruthenates with edge-sharing bonds}

\begin{figure*}[t]
	\centering
	\includegraphics[width=0.7\linewidth,trim=0 780 600 0,clip]{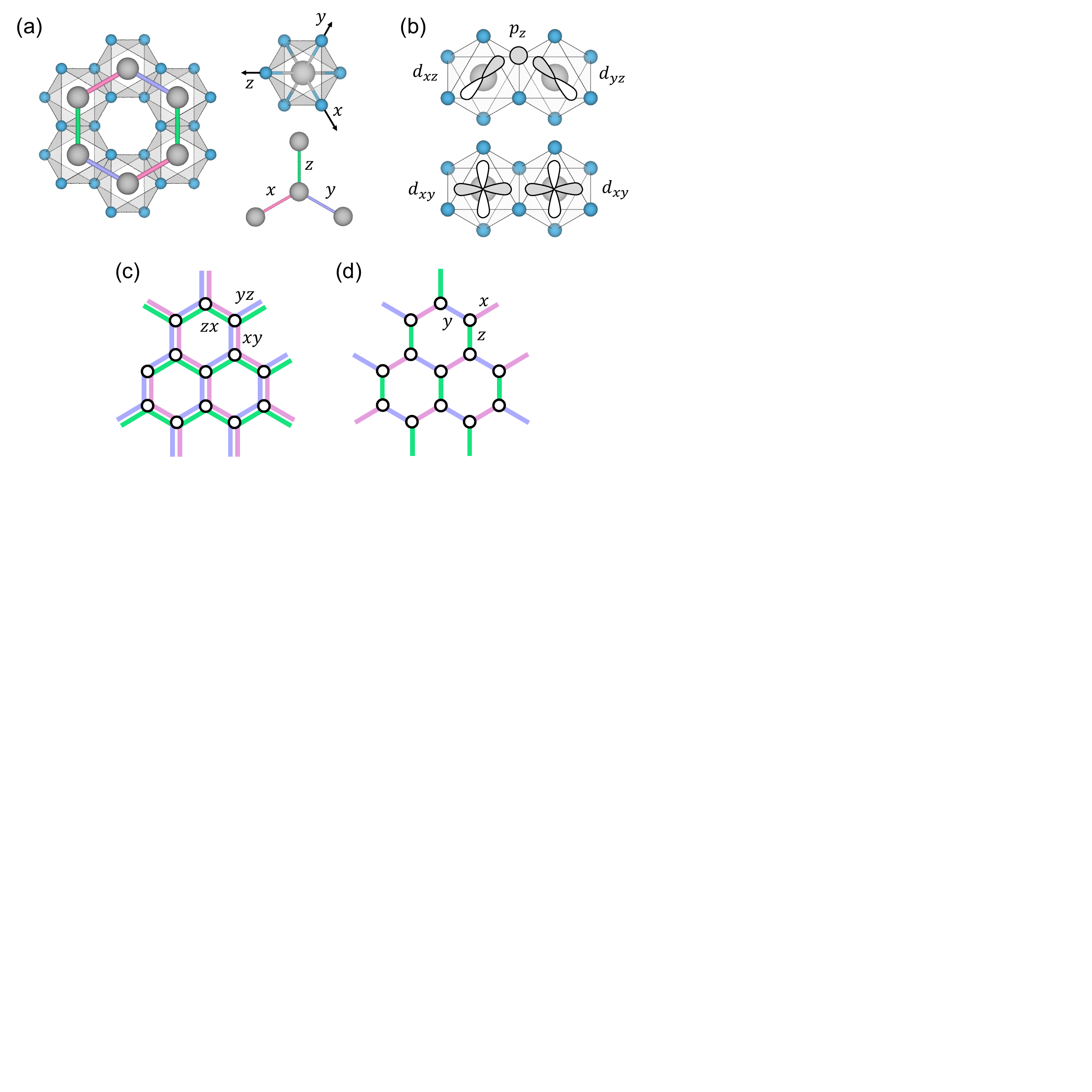}
	\caption{\textbf{Anisotropic triplon hopping on honeycomb lattice.} (a) Honeycomb lattice of edge-sharing RuO$_6$ octahedra. Grey and blue spheres are Ru and O atoms, respectively. Local $x$, $y$ and $z$ directions pointing along Ru--O bonds (top right) as well as $x$-, $y$- and $z$-type bonds (bottom right) are shown. (b) Electron hopping in $\SI{90}{\degree}$ geometry with edge-sharing metal-oxide octahedra, where $\pi$-type overlap (upper) and $\sigma$-type overlap (lower) feature $d$-$p$-$d$ and $d$-$d$ hopping, respectively. (c) Honeycomb pattern for dominant indirect $d$-$p$-$d$ hopping featuring two triplons flavour per bond. (d) Honeycomb pattern for dominant direct $d$-$d$ hopping featuring one triplon flavour per bond. (c) and (d) are reproduced from Ref.~\cite{Takayama2021}.}
	\label{fig:honeycomb}
\end{figure*}

The layered perovskite Ca$_2$RuO$_4$ was thus identified to host excitonic magnetism associated with the spin-orbit-entangled states. A unique feature expected for spin-orbit coupled excitonic magnetism is the emergence of highly anisotropic, bond-dependent magnetic interactions. Such anisotropic coupling and resultant frustrated excitonic magnetism have been proposed to manifest in edge-shared honeycomb lattice of $d^4$ ions~\cite{Khaliullin2013, Chaloupka2019}, as illustrated in Fig.~\ref{fig:honeycomb}(a). In an edge-shared $\SI{90}{\degree}$ bond of RuO$_6$ octahedra, two different exchange processes are present: oxygen-mediated $d$-$p$-$d$ hopping and direct $d$-$d$ hopping of $t_{2g}$ electrons across the octahedron edge (Fig.~\ref{fig:honeycomb}(b)). These two hopping processes provide distinct dynamics of triplon $T_i$. For the $d$-$p$-$d$ hopping process, two triplon components are involved on each bond; $T_x$ and $T_y$ triplons can move along the $z$-bond. The magnetic exchange thus takes the form of bond-dependent $XY$-interaction. On the other hand, in the direct $d$-$d$ hopping process, one triplon is active on each bond, that is, $T_z$ triplon on the $z$-bond. This magnetic coupling is viewed as bond-dependent Ising-type, namely, Kitaev-type interaction.

Such bond-dependent triplon dynamics gives rise to highly frustrated magnetic ground states on a honeycomb lattice. For the dominant $d$-$p$-$d$ hopping, each triplon can travel only along zigzag chain segments of a honeycomb lattice (Fig.~\ref{fig:honeycomb}(c)). The magnetic correlations can develop only along the 1D zigzag chains, which may suppress long-range order and form a spin-nematic state~\cite{Khaliullin2013}. When the direct $d$-$d$ hopping is predominant, the Kitaev-type triplon interaction yields strong frustration and inhibits long-range magnetic order, leading to a bosonic analogue of Kitaev spin liquid~\cite{Anisimov2019} (Fig.~\ref{fig:honeycomb}(d)). The coexistence of both hopping processes in a comparable magnitude will lead to long-range magnetic order. Honeycomb excitonic magnets, namely, edge-shared honeycomb ruthenates, thus appear to be a platform for frustrated excitonic magnetism.

\begin{figure*}
	\centering
	\includegraphics[width=1\linewidth,trim=100 260 80 0,clip]{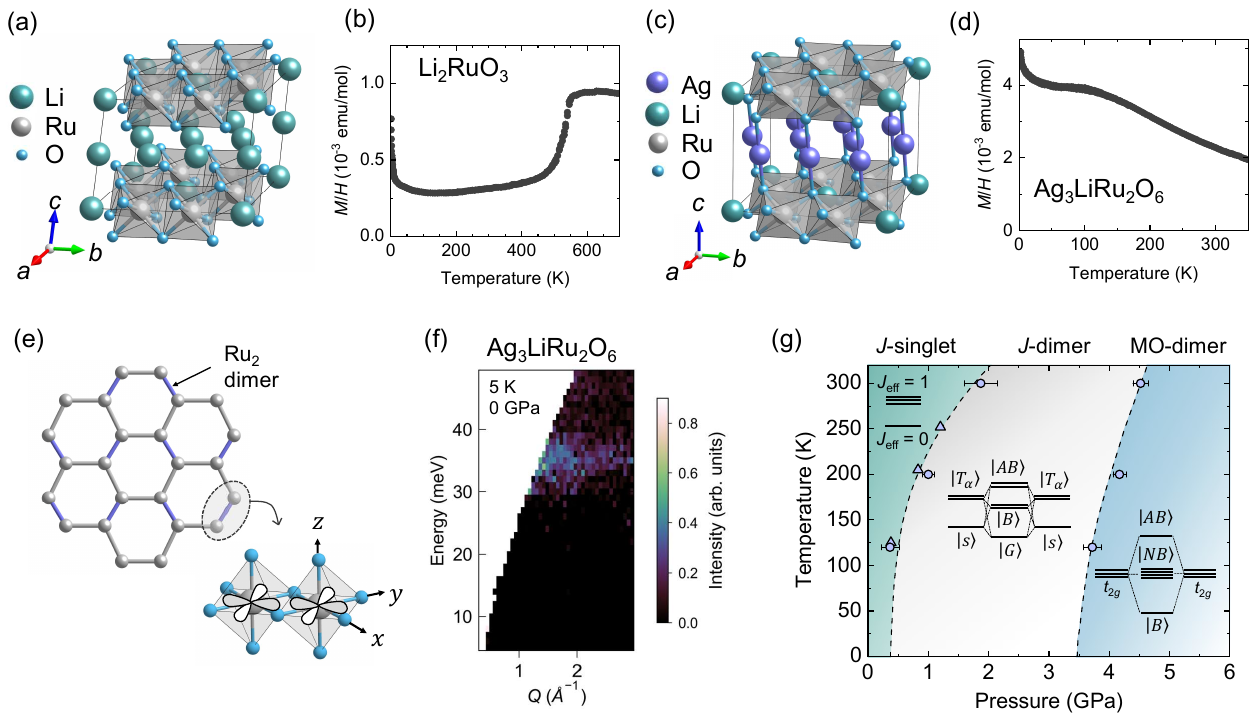}
	\caption{\textbf{Competing spin-orbital singlet phases in honeycomb ruthenates.} (a) Crystal structure of honeycomb Li$_2$RuO$_3$ at $\SI{600}{\kelvin}$~\cite{Miura2009} ($C2/m$, no. 12). Teal, grey and blue spheres are Li, Ru and O atoms, respectively. (b) Temperature-dependent magnetic susceptibility of Li$_2$RuO$_3$, where a structural transition to a non-magnetic valence bond solid state takes place at $\sim\SI{540}{\kelvin}$. Reproduced with permission from Ref.~\cite{Miura2007} ($\copyright$2007 The Physical Society of Japan). (c) Crystal structure of honeycomb Ag$_3$LiRu$_2$O$_6$~\cite{Bette2019} ($C2/m$, no. 12). Teal, purple, grey and blue spheres are Li, Ag, Ru and O atoms, respectively. (d) Temperature dependence of the magnetic susceptibility of Ag$_3$LiRu$_2$O$_6$ with a plateau at around $\SI{110}{\kelvin}$ characteristic of a van Vleck magnetic response. (e) Herringbone pattern of Ru$_2$ dimers (shown in violet) in Li$_2$RuO$_3$ at $\SI{300}{\kelvin}$~\cite{Miura2009} ($P2_1 / m$, no. 11). The direct $d_{xy}$-$d_{xy}$ orbital overlap across the RuO$_6$ octahedra driving molecular orbital formation is shown, which leads to molecular orbital scheme shown in the inset of the MO phase in (g). (f) Inelastic neutron scattering spectrum of Ag$_3$LiRu$_2$O$_6$ at $T = \SI{5}{\kelvin}$ featuring a local excitation from the ground state singlet to the $J_{\rm{eff}} = 1$-derived excited state at $\sim \SI{35}{\meV}$. (g) Pressure-temperature phase diagram of Ag$_3$LiRu$_2$O$_6$. The triangles and circles were determined by magnetisation and neutron diffraction, respectively. $J$-singlet, $J$-dimer and MO-dimer correspond to $J_{\rm{eff}} = 0$ singlet, spin-orbit-entangled dimer and ordinary molecular orbital dimer state. The insets show the $J$-dimer model featuring the ground state $\ket{G}$, the bonding $\ket{B}$ and the antibonding $\ket{AB}$ states originating from the spin-orbit singlet $\ket{s}$ and triplet-derived-doublet $\ket{T_{\alpha}}$. The MO-dimer diagram features the bonding $\ket{B}$, the non-bonding $\ket{NB}$ and the antibonding states $\ket{AB}$ formed from $t_{2g}$ orbitals relevant to the dimers found in Li$_2$RuO$_3$ as well. (d, f, g) are reproduced from Ref.~\cite{Takayama2022}.}
	\label{fig:honeycomb_ruthenates}
\end{figure*}

In real honeycomb ruthenates, however, another competing electronic phase often takes the place of frustrated excitonic magnetism as a consequence of enhanced electron hopping. Honeycomb ruthenates have reduced Ru--Ru separation of $\sim\SI{3}{\angstrom}$, compared to $\sim\SI{3.8}{\angstrom}$ in Ca$_2$RuO$_4$, which facilitates strong direct orbital overlap across the octahedra edges (Fig.~\ref{fig:honeycomb}(b)), leading to a valence bond solid state composed of Ru--Ru molecular orbitals. A honeycomb ruthenate Li$_2$RuO$_3$ (Fig.~\ref{fig:honeycomb_ruthenates}(a)) undergoes a structural transition upon cooling across $\sim\SI{540}{\kelvin}$, below which the magnetic susceptibility becomes small and nearly temperature independent~\cite{Miura2007,Miura2009}, see Fig.~\ref{fig:honeycomb_ruthenates}(b). The transition is characterized by a pronounced Ru--Ru bond disproportionation in which one of three bonds at each Ru contract by $\SI{13}{\percent}$ ($\sim\SI{2.6}{\angstrom}$, shorter even than in Ru metal), forming one dimer per Ru site arranged in a herringbone pattern on the honeycomb lattice, see Fig.~\ref{fig:honeycomb_ruthenates}(e). The eight $t_{2g}$ electrons on each Ru$_2$ dimer fill the bonding and non-bonding molecular orbitals, quenching the local Ru$^{4+}$ moment~\cite{Miura2007}. Li$_2$RuO$_3$ thus demonstrates that another competing electronic phase, valence bond solid, may replace the $J_{\rm{eff}} = 0$ singlet and excitonic magnet.

Chemical substitution has been explored to avoid molecular dimerisation. Substituting Li$^{+}$ with larger Na$^{+}$ ions yields Na$_2$RuO$_3$, where increased Ru--Ru distances ($\sim\SI{3.1}{\angstrom}$) reduce the direct $t_{2g}$ hopping. Indeed, a valence bond solid formation is not observed down to low temperatures; however, the ground state remains debated with conflicting reports pointing to long-range antiferromagnetic order~\cite{Wang2014}, a correlated metallic state~\cite{Veiga2020} or a non-magnetic $J_{\rm{eff}} = 0$ singlet~\cite{Gong2023}.

Alternatively, rigid interlayer bonding can stabilise a nearly regular honeycomb framework against a valence bond solid phase. In Ag$_3$LiRu$_2$O$_6$, linear covalent O--Ag--O dumbbell bonds rigidify the honeycomb lattice and suppress dimerisation, restoring nearly equal Ru--Ru distances ($\sim \SI{3}{\angstrom}$) down to the lowest temperature. Inelastic neutron scattering measurements resolve the local $J_{\rm{eff}} = 0$-like singlet ground state with an excitation at $\sim\SI{35}{\meV}$ to a trigonally-split $J_{\rm{eff}} = 1$-derived doublet, as supported by quantum chemistry calculations~\cite{Takayama2022} (Fig.~\ref{fig:honeycomb_ruthenates}(f)). Although the spin-orbit gap is comparable to that of Ca$_2$RuO$_4$, the magnetic exchange is not strong enough to induce excitonic magnetism, rendering a van Vleck paramagnetic state in Ag$_3$LiRu$_2$O$_6$, see Fig.~\ref{fig:honeycomb_ruthenates}(d). The less pronounced dispersion of the excited states may be partly attributed to the small coordination number of honeycomb lattice and/or frustrated exchange interactions. Application of pressure may enhance $J_{\rm{ex}}$, thus $W_{\rm{t}}$, and induce excitonic magnetism. However, as shown in Fig.~\ref{fig:honeycomb_ruthenates}(g), applying hydrostatic pressure gives rise to other spin-orbital singlet phases. At intermediate pressure above $\SI{0.5}{\GPa}$ and below $\SI{4}{\GPa}$, a spin-orbit-coupled dimer was observed, where weak dimers of $J_{\rm{eff}} = 0$-like states are stabilised by a pseudo Jahn-Teller admixture of  $J_{\rm{eff}} = 1$ states into the singlet. Beyond $\SI{4}{\GPa}$, an ordinary molecular orbital dimer resembling that of Li$_2$RuO$_3$ resurges. So far, the attempts to tune exchange in honeycomb ruthenates are have been consistently pre-empted by valence bond instability. To realise frustrated excitonic magnetism on a honeycomb lattice, further materials development is required where reduction of effective spin-orbit gap by crystal field, enhanced exchange interaction and avoided valence bond solid formation need to be incorporated.

\subsubsection*{$A_2$Ru$_2$O$_7$: corner-sharing pyrochlores with an intermediate bond angle}

\begin{figure*}
	\centering
	\includegraphics[width=0.75\linewidth,trim=100 270 270 0,clip]{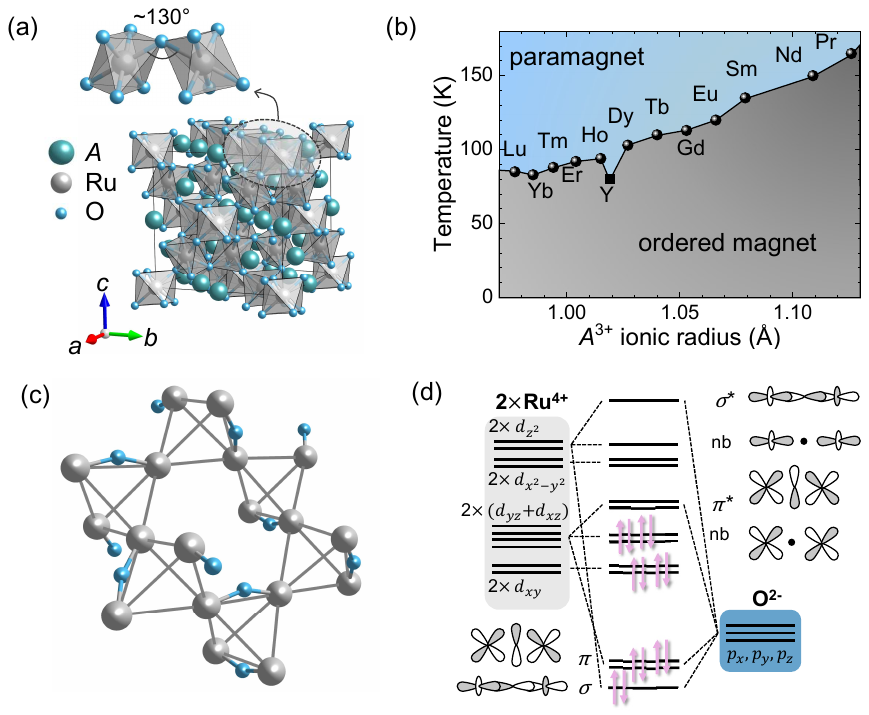}
	\caption{\textbf{Excitonic magnetism and valence bond solid in pyrochlore ruthenates.} (a) Crystal structure of pyrochlore $A_2$Ru$_2$O$_7$ where $A$ is a trivalent cation ($Fd\overline{3}m$, no. 227). Teal, grey and blue spheres are $A$, Ru and O atoms, respectively. The upper figure highlights local connectivity of RuO$_6$ octahedra with an intermediate Ru--O--Ru angle of $\sim \SI{130}{\degree}$. (b) Phase diagram of pyrochlore ruthenates as a function of $A^{3+}$ cation radius based on magnetisation and thermodynamic measurements from Ref.~\cite{Ito2001}. Ordering temperature $T_{\rm{N}}$ falls with decreasing $A^{3+}$ cation radius. (c) Pattern of Ru$_2$O orbital molecules observed in In$_2$Ru$_2$O$_7$. Grey and blue spheres correspond to Ru and O atoms, respectively. Ru$_2$O molecules beyond the range of the shown fragment are terminated at the O atom. (d) Molecular orbital (MO) scheme of Ru$_2$O unit with $\SI{180}{\degree}$ Ru--O--Ru geometry. $\sigma$,$\pi$ and $\sigma^{\ast}$, $\pi^{\ast}$ denote bonding and antibonding MO, while nb corresponds to non-bonding MO. (c) and (d) reproduced with permission from Ref.~\cite{Krajewska2024} ($\copyright$2024 American Association for the Advancement of Science).}
	\label{fig:pyrochlore_ruthenates}
\end{figure*}

To reach intermediate exchange strengths, below those in the $\SI{180}{\degree}$ perovskite limit but greater than those in the $\SI{90}{\degree}$ edge-sharing honeycomb, the corner-sharing pyrochlore ruthenates $A_2$Ru$_2$O$_7$ ($A$: trivalent cation) provide a suitable platform. The Ru$^{4+}$ cations form a geometrically frustrated network of corner-sharing tetrahedra, called the pyrochlore lattice, built from corner-sharing RuO$_6$ octahedra (Fig.~\ref{fig:pyrochlore_ruthenates}(a)). The RuO$_6$ octahedra are trigonally compressed along the axes pointing towards the centre of the Ru$_4$ tetrahedron. The pyrochlore structure accommodates $A^{3+}$ cations across a wide range of ionic radii by trigonally compressing the RuO$_6$ octahedra and bending the Ru--O--Ru bond with an intermediate angle of $\sim \SI{130}{\degree}$. The largest cation Pr$^{3+}$ gives moderate compression and a bond angle of $\sim\SI{131.5}{\degree}$~\cite{Kennedy1995}, while the smallest Lu$^{3+}$ cation gives enhanced compression and a reduced angle of $\sim\SI{127}{\degree}$~\cite{Bertaut1959}.

The $A_2$Ru$_2$O$_7$ series display long-range antiferromagnetic order of Ru moments with a transition temperature $T_{\rm{N}}$ that drops monotonically with the decreasing size of A$^{3+}$ and hence with the decreasing Ru--O--Ru angle ($A$ = Pr: $T_{\rm{N}} \sim \SI{160}{\kelvin}$, $A$ = Lu: $T_{\rm{N}} \sim \SI{75}{\kelvin}$)~\cite{Ito2001,Lee2023}, see Fig.~\ref{fig:pyrochlore_ruthenates}(b). The bond angle, scaled by the $A$-site cation, therefore controls the transfer integral and $J_{\rm{ex}}$. The antiferromagnetic order of the Ru sublattice is characterised by a $\mathbf{q} = (0,0,0)$ propagation vector, where two non-collinear magnetic orders indistinguishable by neutron powder diffraction have been proposed~\cite{Ito2000,Ito2001,Taira2003,Gurgul2007,Chang2010,vanDuijn2017,Ku2018,RuizBustos2024,Museur2026,Yadav2026}. Ordered moments of pyrochlore ruthenates are somewhat reduced from the spin-only $S = 1$ value, namely $\sim\SI{1.4}{\bohrmagneton}$ in Y$_2$Ru$_2$O$_7$ and $\sim\SI{1.2}{\bohrmagneton}$ in Nd$_2$Ru$_2$O$_7$~\cite{Ito2000}, which could be a signature of frustration, strong $d$-$p$ hybridisation, or unquenched orbital moment.

The relevance of SOC to the electronic and magnetic properties of $A_2$Ru$_2$O$_7$ has been generally dismissed, until recent experimental and theoretical works anchored them as excitonic magnets. Ru $L_3$-edge RIXS directly resolves a $J_{\rm{eff}} = 0$-like singlet ground state in Y$_2$Ru$_2$O$_7$ with a multiplet structure compatible with a spin-orbit-entangled $d^4$ ion~\cite{Krajewska2024}. A similar picture had, in fact, already been proposed for Y$_2$Ru$_2$O$_7$ from ${}^{99}$Ru Mössbauer spectroscopy~\cite{Kmiec2006}. Recent microscopic theory also reproduces the experimentally observed antiferromagnetic magnetic orders in several pyrochlore ruthenates within the excitonic magnetism framework~\cite{Swarup2026}.

Approaching excitonic QCP may require reducing exchange interactions further, which is scaled by the Ru--O--Ru angle through the $A$-site cation size. In$^{3+}$ is smaller than any rare-earth cation, and In$_2$Ru$_2$O$_7$ stabilised by high pressure synthesis features the most trigonally compressed RuO$_6$ and the smallest Ru--O--Ru angle of $\sim\SI{126}{\degree}$~\cite{Krajewska2024}, placing it closer to the threshold of excitonic QCP. Despite the largest trigonal distortion of RuO$_6$, its single-ion electronic structure is that of a $J_{\rm{eff}} = 0$-like singlet state (Fig.~\ref{fig:RIXS}(b)). Ru $L_3$-edge RIXS resolves intra-$t_{2g}$ excitations with $J_{\rm{eff}} = 1$-derived doublet at $\sim\SI{50}{\meV}$, and the obtained electronic parameters, such as $\lambda_{\rm{SOC}} \sim \SI{70}{\meV}$, ${\Delta}_{\rm{cf}} \sim \SI{0.3}{\eV}$ and $J_{\rm{H}} \sim \SI{0.31}{\eV}$, are similar to those of Ca$_2$RuO$_4$.

However, the expected excitonic magnetism near the QCP regime is not realised in In$_2$Ru$_2$O$_7$. On cooling, it undergoes multiple first-order structural and electronic phase transitions in a narrow temperature window around $\SI{250}{\kelvin}$, with a large release of entropy~\cite{Krajewska2024}. Below $\SI{250}{\kelvin}$, the magnetic susceptibility drops sharply and remains nearly temperature independent, similarly to Li$_2$RuO$_3$. However, instead of metal-metal dimerisation as that found in the honeycomb ruthenates or the pyrochlore Y$_2$Mo$_2$O$_7$~\cite{Thygesen2017}, In$_2$Ru$_2$O$_7$ develops a massive local distortion of Ru-O bonds. One of the six bonds of each RuO$_6$ octahedron shrinks by $\SIrange{10}{13}{\percent}$ down from $\sim\SI{2}{\angstrom}$ and combine into nearly linear, semi-isolated Ru$_2$O units form on top of the pyrochlore network as shown in Fig.~\ref{fig:pyrochlore_ruthenates}(c). The $J_{\rm{eff}} = 0$ Mott insulator is replaced by a molecular-orbital band insulator, where Ru 4$d$ and O 2$p$ electrons fill the bonding and non-bonding molecular orbitals within each Ru$_2$O unit and a charge gap is opened without Coulomb $U$ (Fig.~\ref{fig:pyrochlore_ruthenates}(d)). In$_2$Ru$_2$O$_7$ therefore evades the excitonic quantum critical regime through valence bond solid formation.

This behaviour distinguishes In$_2$Ru$_2$O$_7$ from the rest of the pyrochlore ruthenate series. The rare-earth pyrochlore ruthenates realise the excitonic magnet ground state without a structural transition. This points to a driving force beyond the competition between the effective gap ${\Delta}_{\rm{eff}}$ and $W_{\rm{t}}$. The most likely candidate is the covalent character of In-O bond, in contrast to the ionic $A$-O bonds from the rare-earth members. This covalency drives a pronounced disproportionation of the InO$_8$ polyhedra already above room temperature. On further cooling this disproportionation intensifies and some O atoms are expelled from the InO$_8$ polyhedra, which form the short Ru--O bonds of the Ru$_2$O molecules~\cite{Krajewska2024}. Thus, the instability associated with In--O bonds triggers the structural transitions and pushes In$_2$Ru$_2$O$_7$ away from the excitonic QCP. Interestingly, a similar mechanism of disproportionation of covalent Tl--O bonds has been discussed to drive the metal insulator transition in Tl$_2$Ru$_2$O$_7$~\cite{Ishii2000}, where the non-magnetic ground state is interpreted as $S = 1$ Haldane chain~\cite{Lee2006}.

\section*{Emergent picture for spin-orbit-entangled $d^4$ systems}

\begin{figure*}
	\centering
	\includegraphics[width=0.8\linewidth,trim=100 390 300 0,clip]{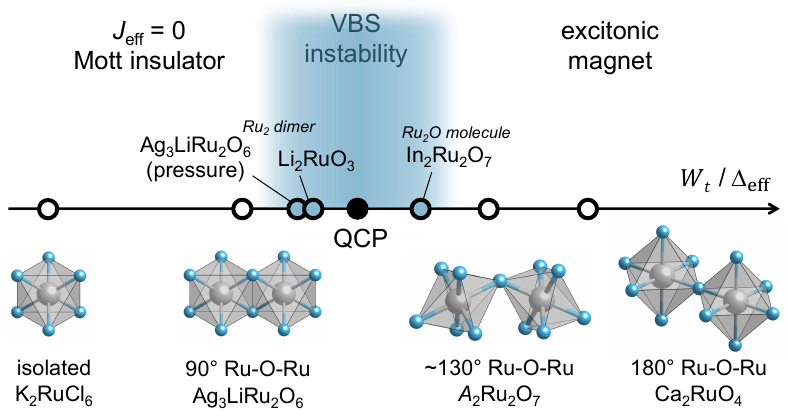}
	\caption{\textbf{Material phase diagram of $d^4$ Mott insulating ruthenium compounds.} The compounds are organized by the ratio between the triplon bandwidth and the effective singlet-triplet gap $W_{\rm{t}}$ / ${\Delta}_{\rm{eff}}$, which can be varied through the connectivity and bonding geometry of Ru-ligand octahedra and their local distortion. K$_2$RuCl$_6$ with isolated RuCl$_6$ octahedra represents the single-ion $J_{\rm{eff}} = 0$ limit, whereas Ca$_2$RuO$_4$ with corner-sharing RuO$_6$ octahedra realises excitonic magnetism. Honeycomb ruthenates with edge-sharing $\sim \SI{90}{\degree}$ Ru--O--Ru bonds lie on the $J_{\rm{eff}} = 0$ side, while rare-earth pyrochlore ruthenates with $\sim \SI{130}{\degree}$ Ru--O--Ru bonds exhibit excitonic magnetic order. Near the anticipated quantum critical point (QCP), competing valence bond solid (VBS) instabilities may intervene, including Ru$_2$ molecular-orbital dimers in Li$_2$RuO$_3$, spin-orbit-entangled and molecular-orbital dimers in Ag$_3$LiRu$_2$O$_6$ under pressure, as well as Ru$_2$O orbital molecules in In$_2$Ru$_2$O$_7$. These instabilities highlight the importance of lattice geometry and chemical bonding beyond the balance of $W_{\rm{t}}$ and ${\Delta}_{\rm{eff}}$.}
	\label{fig:material_phase_diagram}
\end{figure*}

The ruthenium compounds surveyed here span two limiting regimes of spin-orbit entangled $d^4$ physics: K$_2$RuCl$_6$, a $J_{\rm{eff}} = 0$ van Vleck paramagnet with negligible intersite exchange, and Ca$_2$RuO$_4$, an excitonic magnetic with a longitudinal Higgs mode excitation. These two compounds anchor a conceptual phase diagram organised by the ratio $W_{\rm{t}} / {\Delta}_{\rm{eff}}$, where the triplon bandwidth $W_{\rm{t}}$ is governed by exchange interactions, lattice connectivity, and magnetic frustration, while the effective singlet-triplet gap ${\Delta}_{\rm{eff}}$ is determined by spin-orbit coupling and non-cubic crystal fields (Fig.~\ref{fig:material_phase_diagram}).

Within this framework, excitonic quantum criticality is anticipated when $W_{\rm{t}}$ is comparable to ${\Delta}_{\rm{eff}}$. Real materials, however, reveal that this critical regime is often pre-empted by competing singlet instabilities. As the QCP is approached from the $J_{\rm{eff}} = 0$ Mott insulator limit, enhancement of electron hopping to increase $W_{\rm{t}}$ can simultaneously favour molecular-orbital formation and the opening of a bonding-antibonding gap. When the QCP is approached from excitonic magnet regime, where the ordered moment and the associated energy gain diminish, other electronic states may become competitive. Thus, the electronic processes that promote excitonic magnetism may also provide routes towards competing non-magnetic singlet states.

The honeycomb and pyrochlore ruthenates illustrate this competition from opposite sides of the phase diagram. In the honeycomb ruthenates, bond- and flavour-dependent triplon hopping should stabilise exotic states such as spin nematic and bosonic Kitaev spin liquid. The direct $d$--$d$ hopping that generates Kitaev-type triplon coupling also drives Ru$_2$ molecular orbital formation. When the dimerisation is suppressed by interlayer O--Ag--O bonds, as in Ag$_3$LiRu$_2$O$_6$, the system remains $J_{\rm{eff}} = 0$-like singlet, and enhancing exchange by pressure leads first to a spin-orbit-coupled dimer and then ordinary molecular orbital dimer. For the pyrochlore ruthenates, the boundary can be approached from the excitonic side. Rare earth $A_2$Ru$_2$O$_7$ are excitonic magnets whose $J_{\rm{ex}}$ decreases with Ru--O--Ru angle, therefore In$_2$Ru$_2$O$_7$, with the smallest angle, is expected to lie closest to the QCP. Nevertheless, it forms non-magnetic Ru$_2$O orbital molecules, driven not by Ru--Ru dimerisation but by the disproportionation of covalent In--O bonds.

The contrasting role of bonds outside the Ru network is particularly striking. Interlayer covalent O--Ag--O bonds in Ag$_3$LiRu$_2$O$_6$ suppress dimerisation, whereas covalent In--O bonds in In$_2$Ru$_2$O$_7$ promote orbital-molecule formation. The bonding character and the geometry of the spectator ions thus emerges as an additional tuning knob, and the $W_{\rm{t}} / {\Delta}_{\rm{eff}}$ ratio alone does not capture the competition between excitonic magnetism and singlet instabilities. Realising quantum criticality therefore requires not only tuning the balance between $W_{\rm{t}}$ and ${\Delta}_{\rm{eff}}$, but also avoiding lattice-assisted singlet instabilities. One possible strategy is reducing ${\Delta}_{\rm{eff}}$ through crystal-field control, whilst retaining a spin-orbit-entangled state and suppressing bond disproportionation. This remains a key challenge for exploring excitonic criticality and associated unconventional spin-orbital phases. Besides the search for the excitonic QCP, several further directions may extend spin-orbit singlet-triplet physics to unconventional magnetic phases and electronic states, as discussed below.

\section*{Outlook}

The 4$d^4$ materials evaluated here typically feature a positive non-cubic crystal field ${\Delta}_{\rm{cf}}$ that splits the local $J_{\rm{eff}} = 1$ triplet to stabilize a doublet component. It has been proposed that in systems with a large negative non-cubic crystal field via tetragonal or trigonal elongation, a singlet state out of $J_{\rm{eff}} = 1$ triplet is isolated and comes close to the ground state singlet. Inter-site exchange operating on this singlet-singlet manifold maps onto an effective pseuduspin-1/2 model, where their magnetism is described by the transverse-field Ising model~\cite{Chaloupka2024}. The intrinsic transverse field here is set internally by the singlet-singlet gap ${\Delta}_{\rm{eff}}$. Experimental realization of this scenario would provide a physical platform to explore emergent phases such as a clock order and the intermediate Berezinskii-Kosterlitz-Thouless phase~\cite{Li2020}. 

$d^4$ ruthenium compounds often locate in the vicinity of metal-insulator transition, as seen in the progression from insulating Ca$_2$RuO$_4$ to Sr$_2$RuO$_4$~\cite{Maeno1994} or Ca$_3$Ru$_2$O$_7$~\cite{Cao1997}, and in pyrochlore ruthenates where carrier doping can induce a metallic state~\cite{Kaneko2021}. For Ca$_3$Ru$_2$O$_7$, an itinerant magnet, RIXS spectra indicate that the spin-orbit multiplets can persist even in the metallic phase, where the system may be viewed as an itinerant excitonic magnet~\cite{Bertinshaw2021}. A metallic state in close proximity of excitonic criticality has been proposed to host a ferromagnetic state or even unconventional superconductivity~\cite{Chaloupka2016}. Determining how spin-orbit excitons and their collective excitations develop across both filling- and bandwidth-controlled insulator-to-metal transitions represents a pivotal frontier that links $d^4$ excitonic insulators to the broader landscape of correlated spin-orbit-coupled metals.

Lastly, in addition to $d^4$ compounds, $d^8$ electronic configuration in a tetrahedral crystal field, which features a fully occupied lower $e$ doublet and an upper $t_2$ triplet with $4$ electrons, can be another platform for excitonic magnetism. The upper $t_2$ manifold realises a multiplet structure analogous to $d^4$ ions in octahedral environment, with a non-magnetic $J_{\rm{eff}} = 0$ ground singlet and an excited $J_{\rm{eff}} = 1$ triplet~\cite{Abragam1970,LiChen2019}. Tetrahedral Ni$^{2+}$ spinels are a possible platform~\cite{Das2019}. $\lambda_{\rm{SOC}}$ of Ni$^{2+}$ in octahedral environment is $\sim \SI{40}{\meV}$~\cite{Boca2004}, however, in spinel NiRh$_2$O$_4$ a reduced value of $\SIrange{15}{30}{\meV}$ was reported which has been attributed to covalency effects, while the non-cubic crystal field splitting was found to be $\SIrange{50}{70}{\meV}$~\cite{Chamorro2018,Zager2022}. Tetrahedral $d^8$ systems may be especially promising for realising a small singlet-triplet gap and excitonic magnetism.

\bibliography{bibliography}

\bibliographystyle{npjquantum}

% Following is a new environment, {scilastnote}, that's defined in the
% preamble and that allows authors to add a reference at the end of the
% list that's not signaled in the text; such references are used in
% *Science* for acknowledgments of funding, help, etc.
\newpage

\section*{Acknowledgements} We are grateful to G. Khaliullin, G. Jackeli, A. Yaresko, J. Chaloupka and B. Keimer for fruitful discussions. Images of crystal structures were made with VESTA~\cite{Momma2008}.

\section*{Funding} This work was partly supported by Max Planck Society and Humboldt Foundation.

\section*{Competing interests} The authors declare no competing interests.

\end{document}